\documentclass[aps,prc,twocolumn,showpacs,superscriptaddress]{revtex4-2}  

\usepackage{gensymb}
\usepackage{amssymb}
\usepackage{amsmath}
\usepackage{textcomp}
\usepackage{placeins}

\usepackage{color}
\usepackage{csquotes}
\usepackage{rotating} 
\usepackage{gensymb}
\usepackage{float}
\usepackage{esint}
\usepackage{url}
\usepackage{braket}
\usepackage{threeparttable} 
\usepackage{multirow}
\usepackage{booktabs}
\usepackage{mathtools}  
\usepackage{array}
\usepackage{hyperref}
\usepackage{dirtytalk}
\usepackage{xcolor}
\hypersetup{
    unicode=true,            
    colorlinks=true,         
       citecolor    = blue,
       filecolor = blue,       
       linkcolor = blue,
       urlcolor  = blue,
}

\newcommand{\ra}[1]{\renewcommand{\arraystretch}{#1}}

\makeatletter

\newcommand{\Rmnum}[1]{\expandafter\@slowromancap\romannumeral #1@}
\makeatother

\begin{document}

\title{Stellar \textit{s}-process neutron capture cross sections of $^{69,71}$Ga}



\author{M. Tessler} \email[Corresponding author: ]{moshe.tessler@mail.huji.ac.il}
\affiliation{Soreq Nuclear Research Center, Yavne, Israel 81800}
\affiliation{Racah Institute of Physics, Hebrew University, Jerusalem, Israel 91904}

\author{M. Paul} 
\affiliation{Racah Institute of Physics, Hebrew University, Jerusalem, Israel 91904}

 \author{S. Halfon}
 \affiliation{Soreq Nuclear Research Center, Yavne, Israel 81800}
 
 \author{Y. Kashiv}
 \affiliation{University of Notre Dame, Notre Dame, IN 46556, USA }
 
 \author{D. Kijel}
 \affiliation{Soreq Nuclear Research Center, Yavne, Israel 81800}
 
 \author{A. Kreisel}
 \affiliation{Soreq Nuclear Research Center, Yavne, Israel 81800}

 \author{A. Shor}
 \affiliation{Soreq Nuclear Research Center, Yavne, Israel 81800}
 
 \author{L. Weissman}
 \affiliation{Soreq Nuclear Research Center, Yavne, Israel 81800}

 \date{\today}

 \begin{abstract}
The stable gallium isotopes, $^{69,71}$Ga, are mostly produced by the weak slow ($s$) process in massive stars.
We report here on measurements of astrophysically-relevant neutron capture cross sections of the $^{69,71}$Ga$(n,\gamma)$ reactions.
The experiments were performed by the activation technique using a 
high-intensity ($3-5\times10^{10}$ n/s), quasi-Maxwellian neutron beam that closely mimics conditions of stellar $s$-process nucleosynthesis at $kT \approx$ 40 keV.
The neutron field was produced by a mA proton beam at $E_p=1925$ keV
(beam power of 2-3 kW) from the Soreq Applied Research Accelerator Facility (SARAF),
bombarding the Liquid-Lithium Target (LiLiT).
A 473 mg sample of Ga$_2$O$_3$ of natural isotopic composition was activated in the LiLiT neutron field
and the activities of $^{70,72}$Ga were measured by decay counting via $\gamma$-spectrometry with a high-purity germanium detector.
The Maxwellian-averaged cross sections at $kT$ = 30 keV of $^{69}$Ga and $^{71}$Ga determined in this work are 136(8) mb and 105(7) mb, respectively, in good agreement with previous experimental values.
Astrophysical implications of the measurements are discussed.
\end{abstract}


\maketitle

\section{Introduction}
\begin{figure}[ht]
\centering
  \includegraphics[width=\columnwidth]{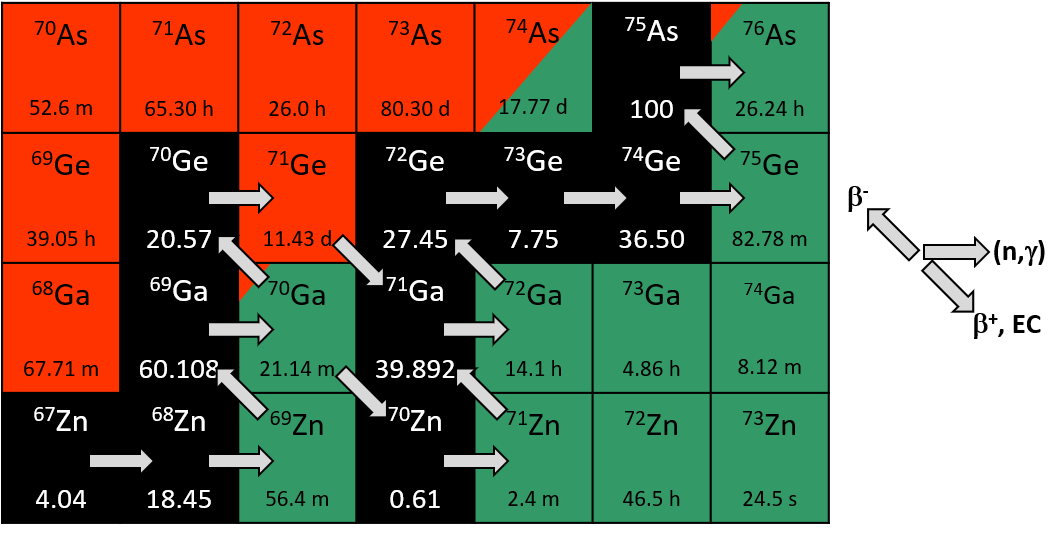}
  \caption{\label{fig:Ga_s_flow}
  The path of the $s$ process between zinc and arsenic.
  The numerical values are the terrestrial isotopic abundances in \% (half-life) for stable (unstable) nuclides.
  When $^{69}$Ga captures a neutron, the product $^{70}$Ga either decays to $^{70}$Ge (99.59\%) or to $^{70}$Zn (0.41\%) with a half-life of 21.14 minutes.
  Following the neutron capture reaction on $^{71}$Ga, the product $^{72}$Ga decays to $^{72}$Ge with a half-life of 14.1 hours.}
\end{figure}

The majority of elements beyond iron are produced by neutron capture reactions in stars.
The slow neutron capture process ($s$ process) is composed of the weak and main components \cite{nucleosynthesis}.
The main component takes place during recurrent thermal pulses in the He shell of low-mass AGB stars ($M  \lesssim 4M_{\odot}$).
The weak component is produced during core He and shell C burning in massive stars ($M  \gtrsim 8M_{\odot}$). 
Neutron densities during the $s$ process are between $10^6$ and $10^{12}$ $n/cm^3$ \cite{REI14}.
The weak $s$ process produces most of the $s$-process isotopes between iron and strontium ($60<A<90$).
Figure \ref{fig:Ga_s_flow} shows the weak $s$-process flow in the Ga region.
Recent simulations show that gallium is the most abundant $s$-element at the end of shell carbon burning in a model of $25M_{\odot}$ star  \cite{PIG10}.

Until recently, there was only a Time-of-Flight (TOF) measurement of the $^{69}$Ga$(n,\gamma)$ cross section,
with a sample of natural gallium \cite{WAL84} (see also \cite{Kadonis}).
Other nuclei measured within the same experimental campaign ($^{74}$Ge, $^{75}$As, $^{81}$Br) show large deviations from  more recent results \cite{Kadonis}.
For $^{71}$Ga, two previous results from activation measurements are in marginal agreement \cite{ANA79, WAL86}.
Very recently, values were published for the neutron capture cross sections of $^{69,71}$Ga \cite{GOB21}
which are stated to disagree with available evaluated data from KADoNiS v0.3 \cite{Kadonis}.
We report here on new measurements of the $^{69,71}$Ga$(n,\gamma)$ cross sections, which were initiated prior to the publication of the latter results.
This work took advantage of the Liquid-Lithium Target (LiLiT) at the Soreq Applied Research Accelerator Facility (SARAF) intense quasi-Maxwellian neutron source (\cite{Paul2019}, see Sec. \ref{saraf_lilit}),
especially valuable for the activation measurement of short-lived nuclides like $^{70}$Ga ($t_{1/2}$= 21.1 min).
Section \ref{saraf_lilit} describes the SARAF-LiLiT neutron source, Sec. \ref{sec:sample} describes the samples used and the irradiation details,
and Sec. \ref{sec:activity} describes the activity measurements. In Sec. \ref{sec:act_results} we report the activation results, in Sec. \ref{sec:cs} we calculate the experimental cross sections,
Sec. \ref{sec:macs} describes how the MACS were calculated, and in Sec. \ref{sec:disc} we discuss the results.

\section{SARAF-LiLiT \label{saraf_lilit}}
\begin{figure}[ht]
\centering
  \includegraphics[width=\columnwidth]{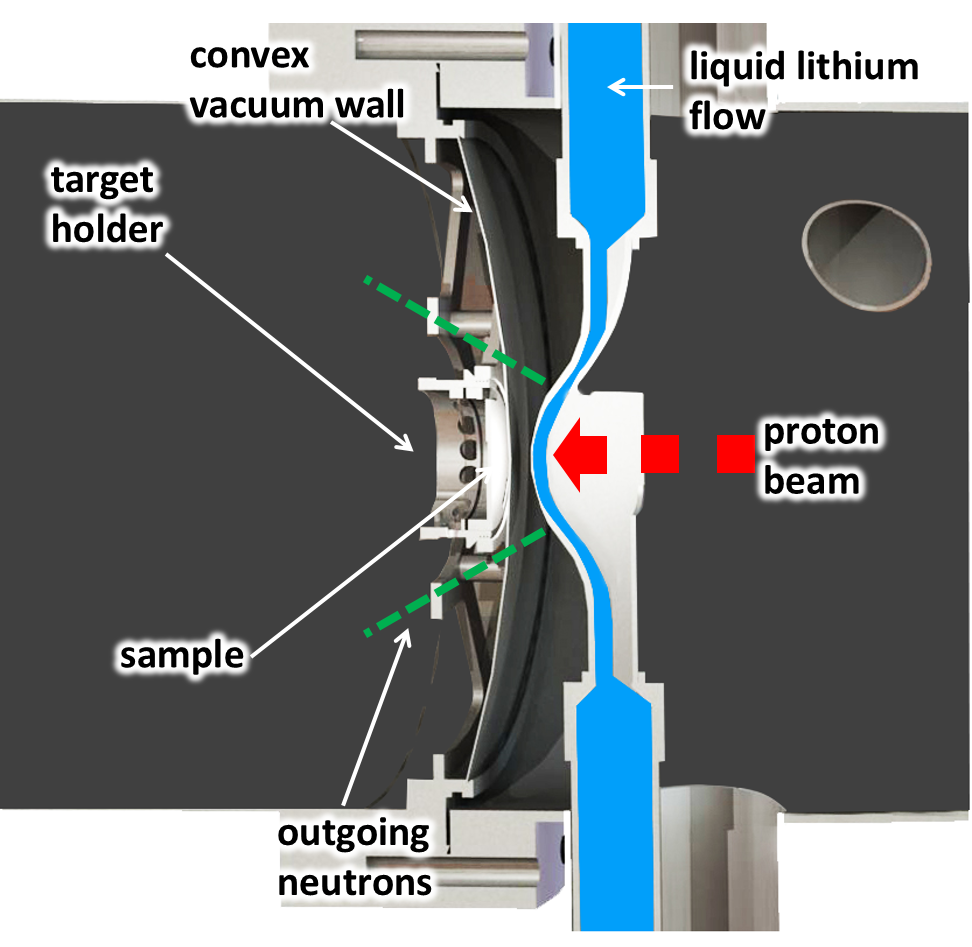}
  \caption{\label{fig:ex_setup} Diagram of the Liquid-Lithium Target (LiLiT) and activation target assembly.
  The 1-2 mA ($\approx$9 mm full width) proton beam (dashed red arrow) impinges directly on the windowless liquid-lithium film.
  The light blue shows the liquid-lithium circulating flow (see \cite{LiLiT1} for details).
  The activation samples are mounted at the center of a ring target holder made of Al and positioned in the outgoing neutron cone (green dashed lines)
  at a distance of $\approx$6 mm from the liquid-lithium film surface.
  The targets are in a vacuum chamber separated from the LiLiT chamber by a 0.5 mm stainless steel concave vacuum wall.}
\end{figure}

An intense $^{7}$Li$(p,n)^{7}$Be neutron source, in the form of a Liquid-Lithium Target (LiLiT) \cite{LiLiT1, LiLiT2},
bombarded by a mA proton beam from the Soreq Applied Research Accelerator Facility (SARAF), was developed and
is used for Maxwellian Average Cross Section (MACS) measurements.
The SARAF accelerator is based on a continuous wave (CW), proton/deuteron RF superconducting linear accelerator.
SARAF Phase I, presently undergoing a major upgrade to its Phase II, consisted of a 20 keV/u Electron Cyclotron Resonance (ECR) ion source injector,
a Low-Energy Beam Transport section (LEBT),
a four-rod Radio Frequency Quadrupole (RFQ, 1.5 MeV/u), a Medium Energy Beam Transport section (MEBT),
a Prototype Superconducting Module (PSM) housing six half-wave resonators and three superconducting solenoids, and a diagnostic plate (D-plate).
The beamline downstream of the accelerator transports the high intensity beam to the target.
SARAF Phase I delivered for experiments currents of up to 2 mA of protons or
deuterons, with energies of up to $\approx$ 4 and 5 MeV, respectively \cite{SARAF1, SARAF2, Mardor2018}.

The SARAF high-intensity beam requires a Li target that can withstand its power,
which is incompatible with solid Li and Li compounds.
The Liquid-Lithium Target (LiLiT) \cite{LiLiT1, LiLiT2} consists of a liquid-lithium film (temperature $\approx$200 \degree C, above the lithium melting 
temperature of 180.5 \degree C) circulated at high velocity (3-7 $m/s$) onto a thin convex stainless-steel support wall.
The target is bombarded with a high-intensity proton beam impinging directly on the Li-vacuum (windowless) interface at an energy above and
close to the $^{7}$Li$(p,n)$ reaction threshold, $E_{th} = 1.88$ MeV (Fig. \ref{fig:ex_setup}).
A rectangular-shaped nozzle just upstream of the curved support wall determines the film width and thickness
to be 18 mm and 1.5 mm, respectively (see \cite{LiLiT1} for details).
The first few microns at the surface of the liquid-lithium film serve as the neutron-producing thick target.
The deeper Li film layers act as a beam dump, from which the power is transported by the flow to a heat exchanger.
A spherical cap made of stainless steel foil, 0.5 mm thick and 19 cm in diameter,
is located $\approx$1 mm beyond the nozzle and seals the LiLiT vacuum
chamber neutron exit port (Fig. \ref{fig:ex_setup}).
The vacuum wall curvature (convex toward
the Li flow with a curvature radius of 300 mm) allows us to locate a secondary
activation target very close (see below) to the neutron source at the Li-vacuum interface.

\section{Sample characteristics and irradiation details \label{sec:sample}}
In the experiment described here, a 13 mm diameter $^{nat}$Ga$_2$O$_3$
(99.99+\% purity, \cite{Ga2O3}) pellet target was activated.
It was sandwiched between two Au foils (Table \ref{table: samples_Ga}) which were used as monitors of the neutron fluence.
The distance of the 13 mm diameter Ga target from the Li neutron source was $6 \pm 1$ mm, intercepting $> 75\%$ of the outgoing neutrons.
The characteristics of the samples used are summarized in Table \ref{table: samples_Ga}.

The Ga and Au targets were inserted into the LiLiT activation chamber (Fig. \ref{fig:ex_setup}) and held in place by the target holder.
The 9 mm full width proton beam impinged on the free-surface lithium film, resulting in an outgoing neutron cone due to the $^7$Li$(p,n)$ reaction, which irradiated the target.
The setup is shown in Fig. \ref{fig:ex_setup} and explained in the caption.

The number of stable nuclei $A$ per cm$^2$, $n_t(A)$, for a target element with an atomic or molecular mass $M_A$, target area $S$ and mass $m$,
is given by Eq. (\ref{eq: n_t}).
\begin{equation}
 n_t(A) = s_A\cdot a(A) \frac{m \cdot N_A}{S\cdot M_A} \label{eq: n_t}.
\end{equation}
The symbol $N_A$ denotes Avogadro's number and $a(A)$ the isotopic abundance of $A$ (Table \ref{table: decay_Ga}).
The stoichiometry of element $A$ in the target is denoted by $s_A$ ($s_A$=2 or 1 for Ga or Au, respectively).
\begin{table}[htbp]
\centering
\caption{\label{table: samples_Ga}Characteristics of the samples used in this work. They are listed in the order that they were placed downstream from the Li target.
The first gold foil (Au \#33) was necessary to monitor the beam offset (see below).
The Ga oxide targets  \#2 (\#1) were used for neutron irradiations above (below) the Li neutron emission threshold.}
\begin{ruledtabular}
\ra{1.3}
\begin{tabular}{l c c c c}
Sample & Diam. & Mass & Nucleus & $n_t$ \\
 & (mm) & (mg) & & ($10^{19}$cm$^{-2}$)\\[0.5ex]
\hline
Au \#33 & 25 & 109.8(1) &$^{197}$Au & 6.839(4)\\
Au \#14 & 13 & 31.6(1) &$^{197}$Au & 7.28(1)\\
$^{nat}$Ga$_2$O$_3$ \#2 & 13 & 472.5(1) & $^{69}$Ga & 137.5(1)\\
& & & $^{71}$Ga & 91.25(1) \\
Au \#15 & 13 & 32.5(1) &$^{197}$Au & 7.49(1)\\
\midrule
$^{nat}$Ga$_2$O$_3$ \#1 & 13 & 443.0(1) & $^{69}$Ga & 128.9(1)\\
& & & $^{71}$Ga & 85.55(1) \\[1ex]
\end{tabular}
\end{ruledtabular}
\end{table}

The proton beam energy was measured by a TOF pick-up and Rutherford backscattering off a Au target
located in the diagnostic D-plate.
The beam energy was 1925 keV, with an energy spread of $\approx$15 keV.
The energy spread was estimated from beam dynamics calculations and was verified 
experimentally under similar conditions \cite{gitai_thesis}.

\begin{figure}[htbp]
\centering
 \includegraphics[width=0.7\columnwidth]{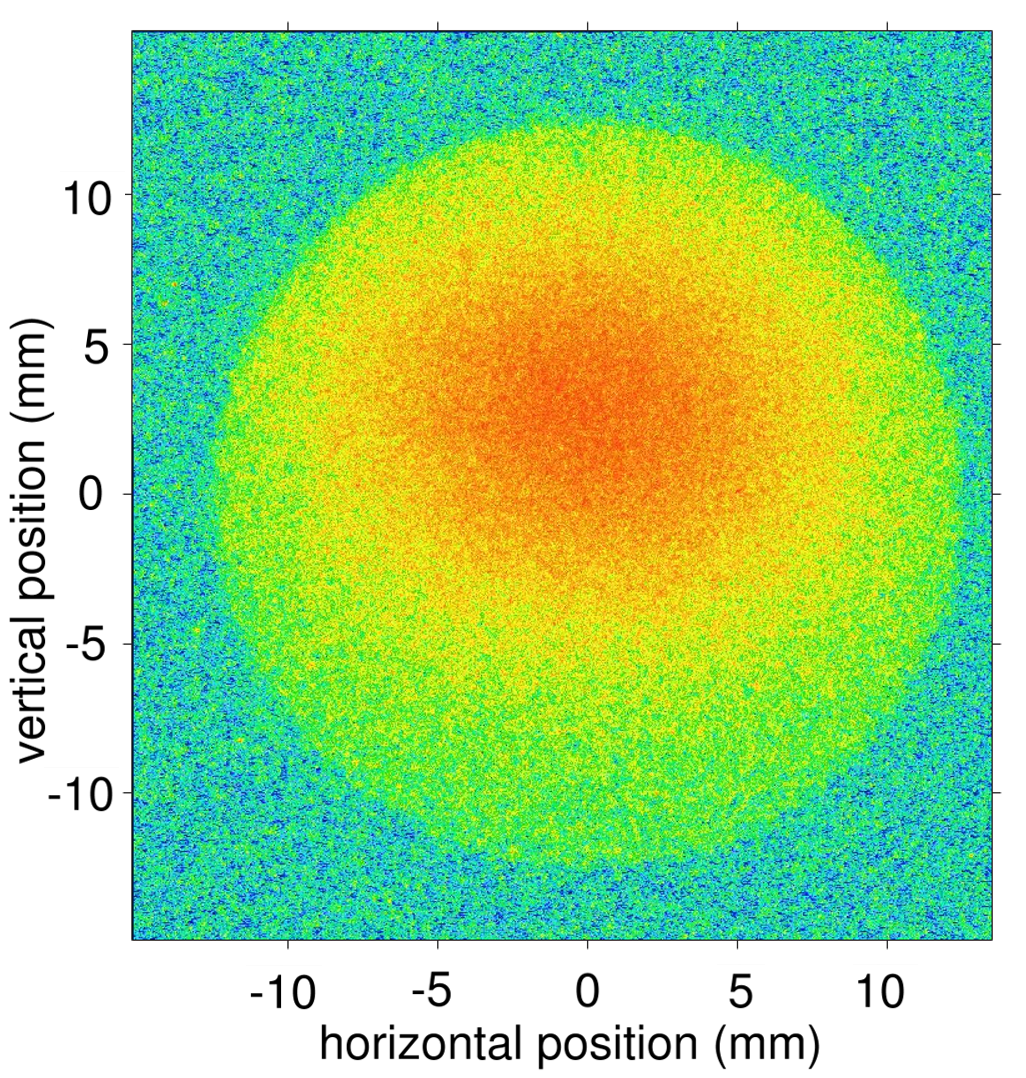}
 \caption{\label{fig: Au_radiography} An auto-radiographic scan of the 25 mm diameter gold foil (Au \#33).
 Red indicates the area with the highest neutron irradiation followed by yellow, green and blue.
 A $\approx$2.5 mm offset was observed in the neutron irradiation and is attributed to the vertical steering of the proton beam.
 This offset was taken into account in the neutron irradiation simulations.}
 \end{figure}

To determine the position of the proton beam relative to the activation targets, the 25 mm diameter gold foil (Au \#33) was auto-radiographically scanned after the samples irradiation (Fig. \ref{fig: Au_radiography}).
An offset of 2.5 mm in the vertical direction was found and accounted for in our detailed simulations.

\begin{figure}[htbp]
\centering
 \includegraphics[width=\columnwidth]{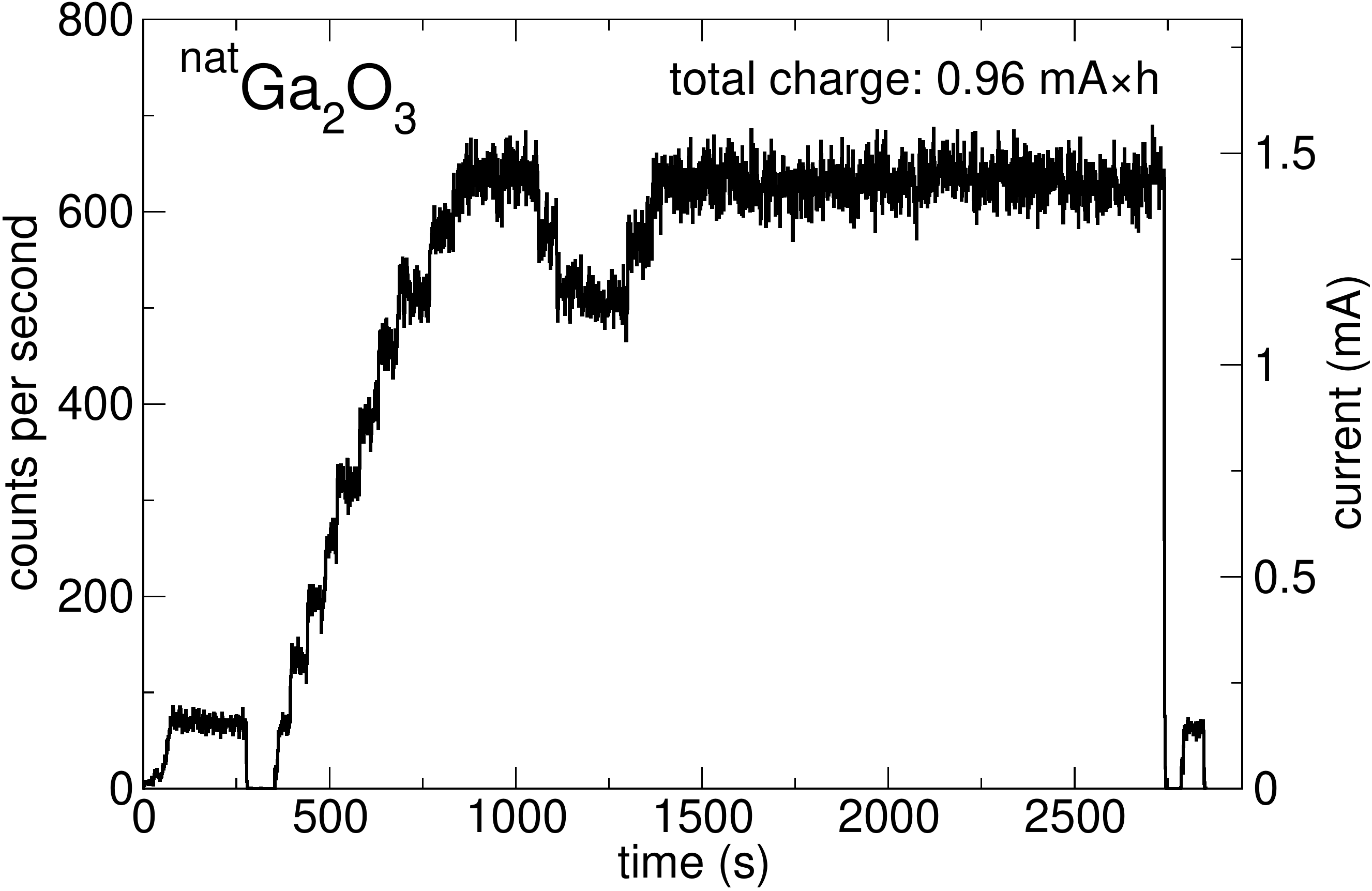}
 \caption{\label{fig: current_Ga}Time record of the fission chamber count rate (left axis)
 and corresponding intensity of the proton beam current (right axis) during
the irradiation.
The two low-intensity intervals and gaps at beginning
and end of the irradiation correspond to the calibration of the fission
count rate against beam current measured at low intensity with a Faraday
cup (see text).
All intensity variations are taken into account by the
$1/f_b$ correction factor (Eq. (\ref{eq:N_act_exp})). The total integrated current was 0.96 mA$\times$h.}
 \end{figure}

Throughout the irradiation, the neutron yield was continuously monitored and recorded with a fission-product ionization
chamber detector (PFC16A, Centronics Ltd.),
counting neutron-induced fission events from a thin $^{235}$U internal foil (1 mg/cm$^{2}$, 12.5 cm$^{2}$ active area).
The fission chamber was located at 0$\degree$ to the incident proton beam, at a distance of $\approx$80 cm downstream from the target.
The fission chamber was covered with a 1 mm thick Cd sheet to absorb scattered thermal neutrons.
The count rate of the fission chamber was calibrated to the beam current at low intensity (10\% duty cycle, using a slow chopper),
with the Faraday cup located $\approx$1 m upstream of the Li target.
After the SARAF was tuned, the beam duty cycle was ramped up to 99\%.
Normally, the ramp up is performed rather slowly while monitoring the temperature and radiation along the beamline and LiLiT.
If necessary, fine-tuning of the beamline ion-optical magnetic elements (bending magnets, steerers)
is performed based on temperature reading of sensors located on the lithium nozzle (see \cite{LiLiT1, LiLiT2} for details).
It is important to know the time dependence of the neutron yield in the case of a short-lived activation product (\textit{e.g.,} $^{70}$Ga, $t_{1/2}$= 21.1 min).
In such a case, one needs to account for fluctuations of the neutron yield when evaluating the fraction of the reaction product that decayed during the irradiation.
The time record of the proton beam current is presented in figure \ref{fig: current_Ga}.
The total integrated current in the irradiation was $\sim$0.96 mA$\times$h.
\FloatBarrier

\section{Activity measurement \label{sec:activity}}
\begin{figure}[htpb]
\centering
 \includegraphics[width=\columnwidth]{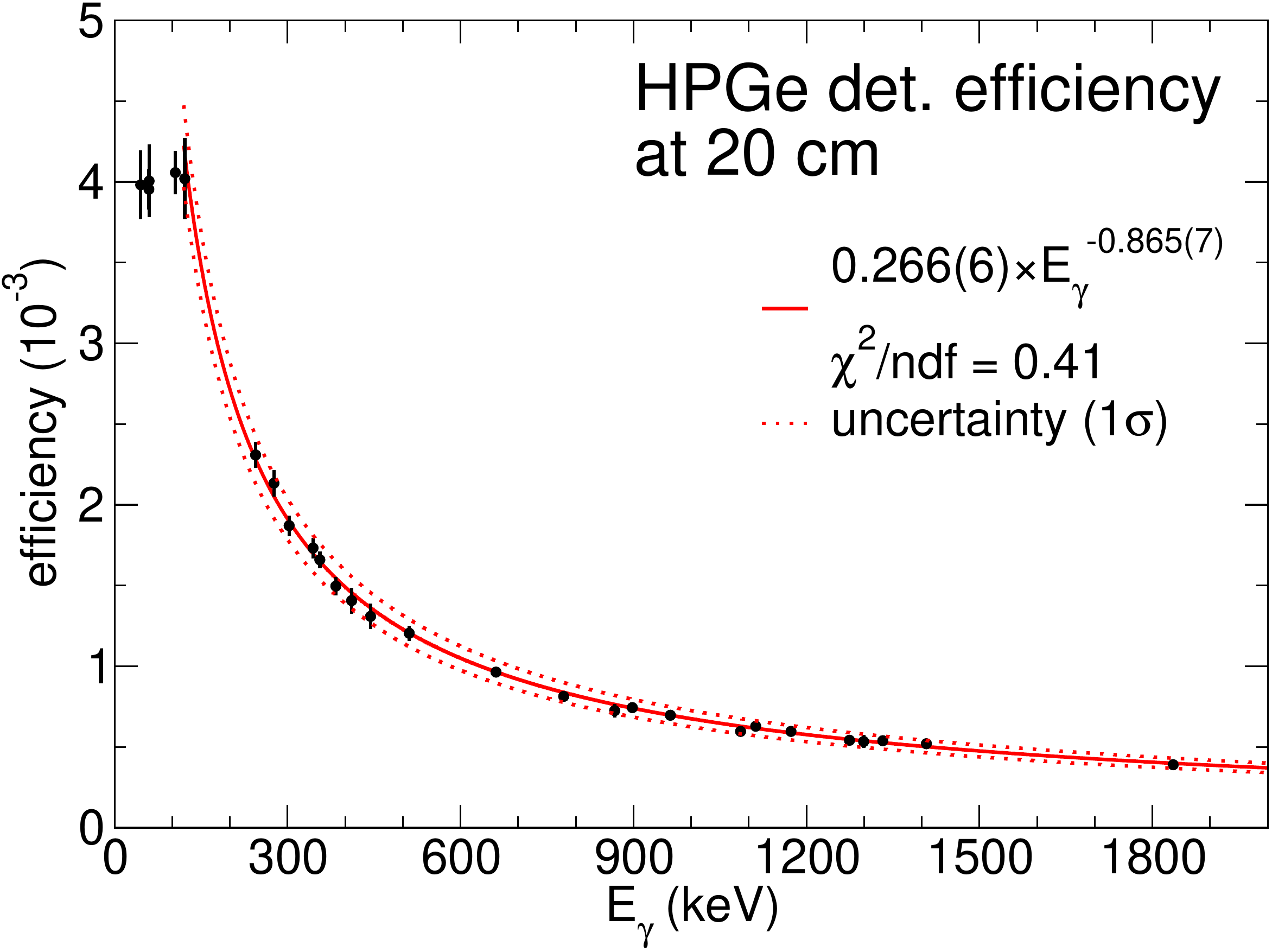}
 \caption{\label{fig: gamma_eff} Efficiency of the HPGe detector used at a distance of 20 cm (black dots with 1$\sigma$ uncertainties) determined by
 standard calibrated radioactive sources: 
 $^{22}$Na, $^{60}$Co, $^{88}$Y, $^{133}$Ba, $^{137}$Cs, $^{241}$Am, $^{152}$Eu and $^{155}$Eu.
 The red curve is a fit to the data with the expression: $\epsilon = a\times E_{\gamma}^{-b}$ and the dotted lines are the uncertainty (1$\sigma$).
 The main source of the efficiency uncertainty is the radioactive source activity; these uncertainties are given by the source manufacturers. The small value of the reduced chi square suggests that these uncertainties are overestimated.}
 \end{figure}

After the irradiation, the induced activities were measured separately for each
sample with a shielded HPGe detector (ORTEC GMX 25-83).
The distance of the sample to the
HPGe detector was 20 cm (2 cm) for the measurement above (below) the neutron threshold.
The detector efficiency was determined by standard calibrated radioactive sources: 
$^{22}$Na, $^{60}$Co, $^{88}$Y, $^{133}$Ba, $^{137}$Cs, $^{241}$Am, $^{152}$Eu and $^{155}$Eu.
The measured efficiency curve at 20 cm is presented in Fig. \ref{fig: gamma_eff}.

\begin{figure}[htpb]
\centering
 \includegraphics[width=\columnwidth]{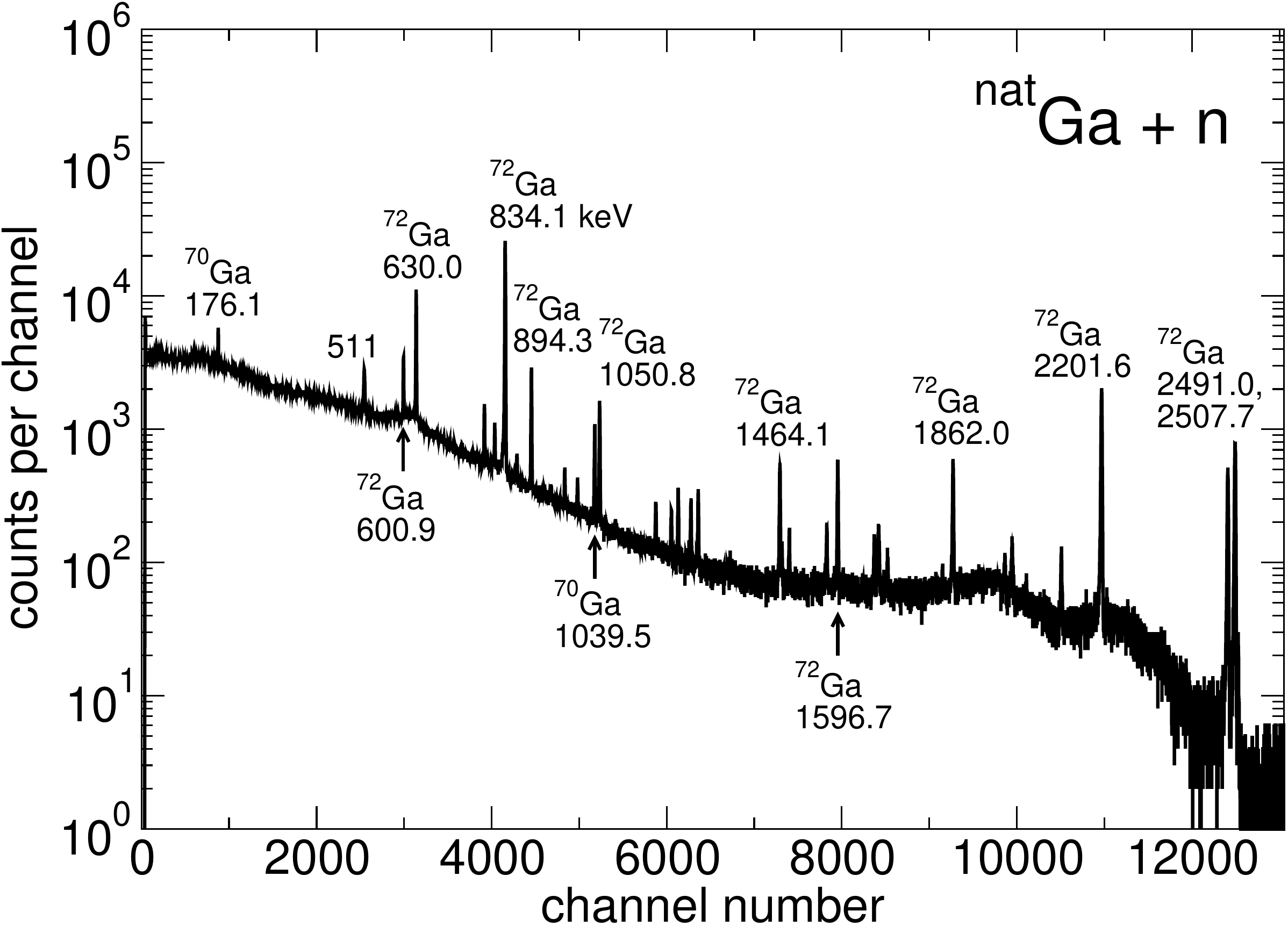}
 \caption{\label{fig: gamma_spec_Ga} $\gamma$-ray spectrum of the activated $^{nat}$Ga$_2$O$_3$ \#2 sample.
 The spectrum was accumulated for 3000 s, starting 2464 s after the end of the neutron activation.
 The sample was located 20 cm from the HPGe detector. The main Ga isotope full-energy peaks, with energies in keV, are labeled.}
 \end{figure}

\begin{figure}[htpb]
\centering
 \includegraphics[width=\columnwidth]{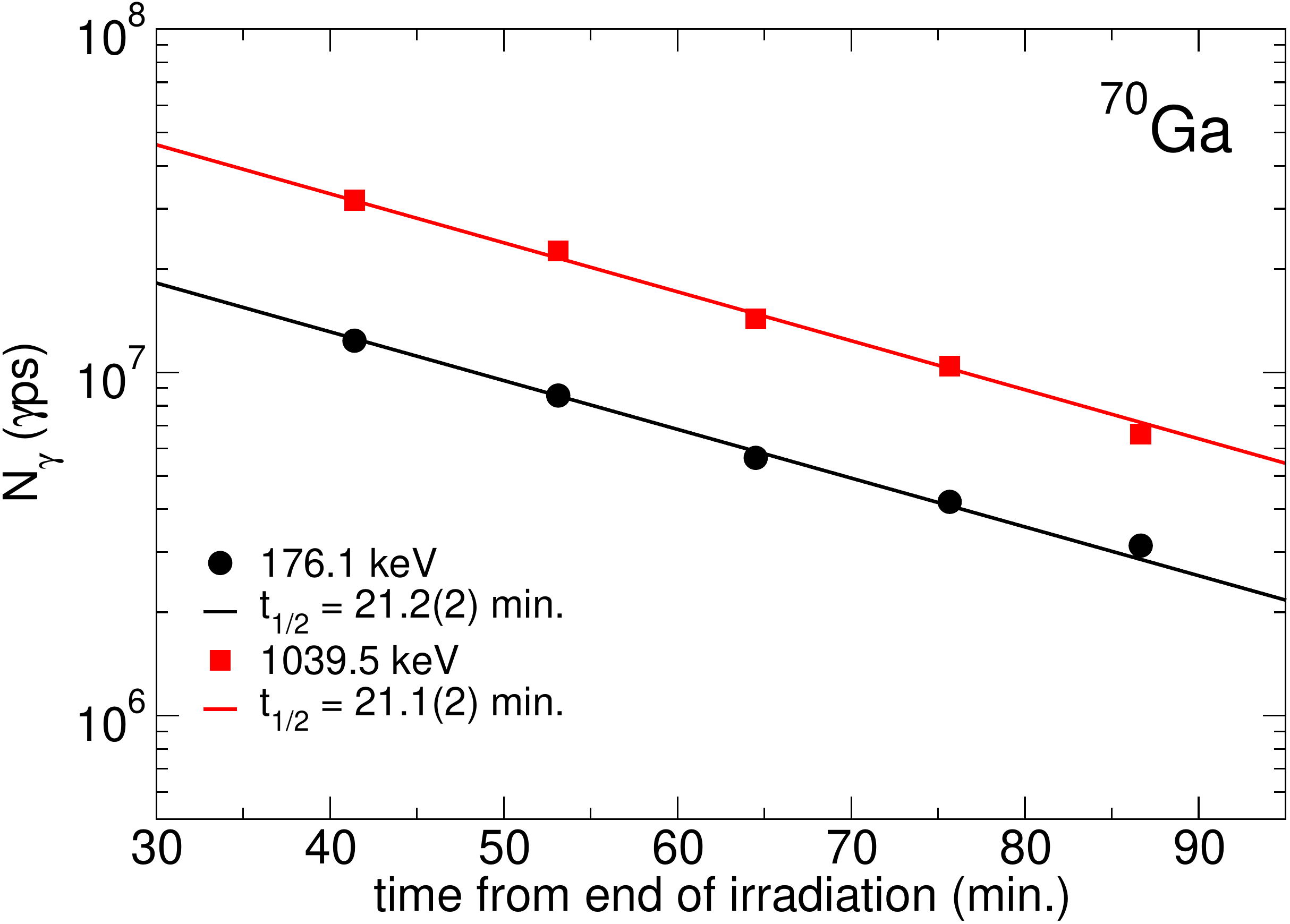}
 \caption{\label{fig:Ga70_decay} Decay curves for $^{70}$Ga. The uncertainties of each data point in this plot are only the statistical uncertainties of the activity measurement.
 $N_{\gamma}= N_{act}{\cdot}I_{\gamma}$ (see Eq. \ref{eq:N_act_exp}) is the photon intensity ($\gamma$s per second)
of the $\gamma$ transitions at the 176.3 and 1039.5 keV lines of the $^{70}$Ga decay, determined by $\gamma$ spectrometry.
Excellent agreement is observed with the adopted half-life of $^{70}$Ga from the literature (21.14(5) min. \cite{GUR16}), and between the number of activated $^{70}$Ga nuclei (see Tables \ref{table:ana_data} and \ref{table: Nact summary_Ga}) derived
independently from each transition, using the respective adopted
$\gamma$ intensities \cite{GUR16}; see text.}
 \end{figure}

The $\gamma$-ray spectrum for the neutron activated Ga sample ($^{nat}$Ga$_2$O$_3$ \#2) is presented in Fig. \ref{fig: gamma_spec_Ga}, where the full-energy peaks of $^{70}$Ga and $^{72}$Ga are labeled.
Figure \ref{fig:Ga70_decay} presents the decay curves of the 176.3 and 1039.5 keV $\gamma$ lines of $^{70}$Ga, showing an excellent agreement with the adopted $^{70}$Ga half-life of 21.14(5) min. \cite{GUR16}.
The numbers of activated $^{70}$Ga nuclei derived from each of the two transitions, using the adopted $\gamma$ intensities \cite{GUR16}, are in excellent agreement as well.
The ratio of $\gamma$ intensities of the 1039 and 176 keV lines
determined here is 2.53(6) in agreement and better precision than the
ratio of adopted intensities 2.24(19) \cite{GUR16}. The reported $\gamma$ intensity ratio in \cite{SCH77} is 2.30(6).
The precise counting of activated $^{70}$Ga nuclei, in spite of its short half-life relative to
the irradiation time and the very low intensity of the $\gamma$ transitions (Table \ref{table: decay_Ga}),
is credited to the high intensity of the LiLiT neutron source.

\begin{figure}[htpb]
\centering
 \includegraphics[width=\columnwidth]{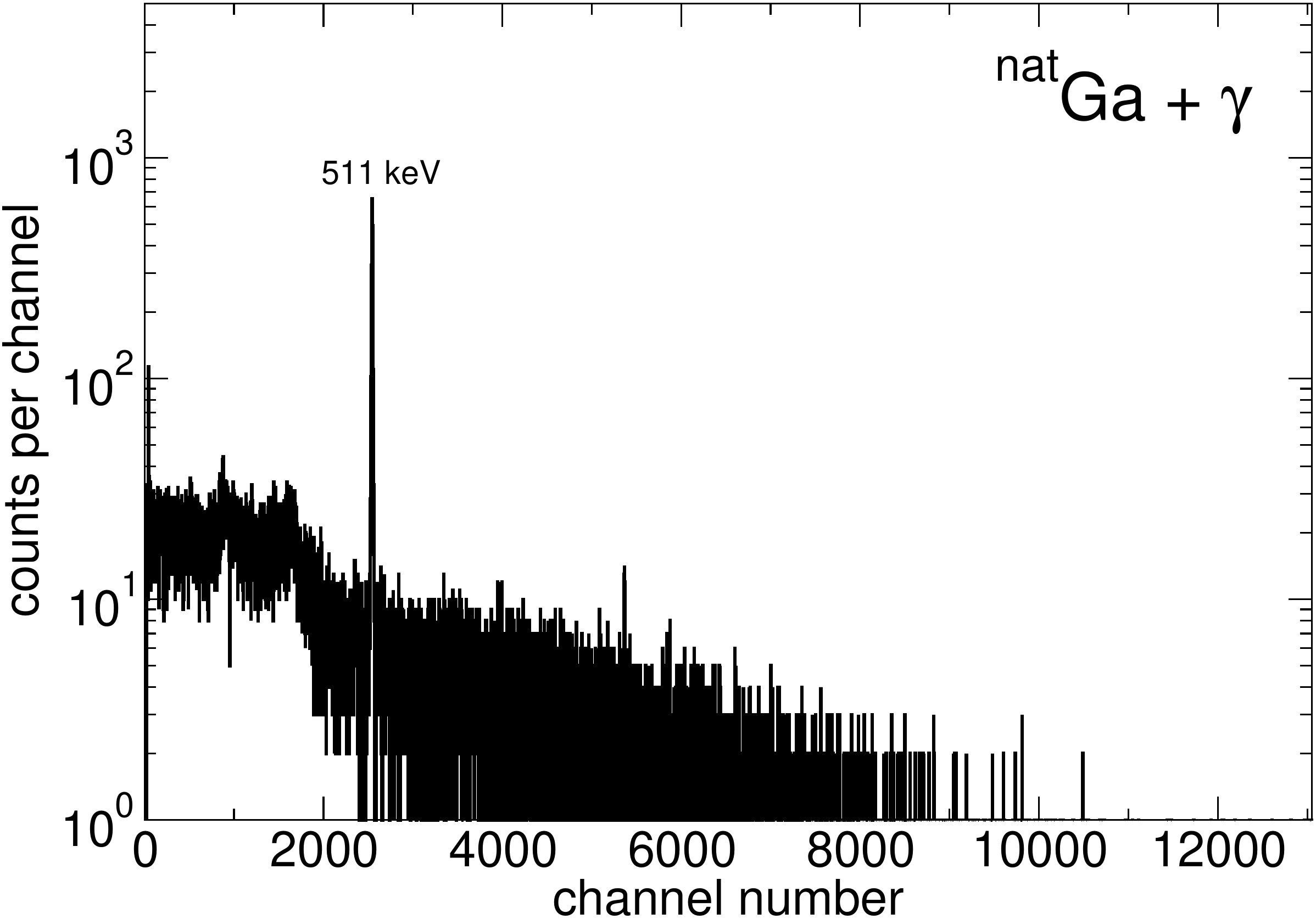}
 \caption{\label{fig: gamma_spec_below_Ga} Secondary $\gamma$-ray spectrum of the $^{nat}$Ga$_2$O$_3$ \#1 sample after irradiation with primary $\gamma$-rays.
 The spectrum was accumulated for 636 s, starting 1275 s after the end of the irradiation.
 The proton energy was below the neutron production threshold, 
 resulting in the production of only $\gamma$ rays.
 The sample was located 2 cm from the HPGe detector.
 No Ga lines are observed.}
 \end{figure}
The SARAF-LiLiT setup was shown to produce high-energy $\gamma$-rays
($\approx$10$^8$ 14.6 and 17.6 MeV $\gamma$-rays/(mA$\times$s)) via the $^7$Li$(p,\gamma)^8$Be reaction \cite{PLB_Zr,Paul2019}.
In cases where there are a stable A-1 isotope, an unstable A isotope and a stable A+1 isotope, the nuclide A can be produced through the A-1$(n,\gamma)$ or A+1$(\gamma,n)$ reactions.
To test the possible contribution of the $^{71}$Ga$(\gamma,n)$ reaction to the production of $^{70}$Ga, an 
irradiation of protons on LiLiT was conducted below the neutron production threshold.
This resulted in the irradiation of the Ga sample ($^{nat}$Ga$_2$O$_3$ \#1) with $\gamma$-rays, but without neutrons.
The proton energy was measured to be 1800 keV.
The total integrated current in this irradiation was $\approx$0.46 mA$\times$h.
The $\gamma$-ray spectrum for the activated Ga sample ($^{nat}$Ga$_2$O$_3$ \#1) is presented in Fig. \ref{fig: gamma_spec_below_Ga}.
As can be seen, no Ga isotopes were created through $(\gamma,n)$ reactions.

\section{Activation results \label{sec:act_results}} 
\begin{table}[ht]
\centering
\caption{\label{table: decay_Ga}Properties of the relevant target and product nuclei studied in this work. The half-life data were taken from \cite{Datasheet198, GUR16, ABR10}.}
\begin{ruledtabular}
\ra{1.3}
\begin{tabular}{l c c c }
Target& Isotopic& Product & Half-life, \\
nucleus& abundance& nucleus &  $t_{\frac{1}{2}}$ \\ [0.5ex]
\hline
$^{197}$Au& 1& $^{198}$Au & $2.6947(3)$ d \\
$^{69}$Ga & 0.60108(9)&$^{70}$Ga & $21.14(5)$ m \\
$^{71}$Ga & 0.39892(9)&$^{72}$Ga & $14.10(1)$ h \\ [1ex]
\end{tabular}
\end{ruledtabular}
\end{table}
The number of activated nuclei created during the irradiation, $N_{act}$, was obtained from the $\gamma$-ray spectra using Eq. (\ref{eq:N_act_exp}),
\begin{equation}
\label{eq:N_act_exp}
 N_{act}=\frac{C}{\epsilon_{\gamma}I_{\gamma}K_{\gamma}}\frac{e^{\lambda t_{cool}}}{1-e^{-\lambda t_{real}}}\frac{t_{real}}{t_{live}}\frac{1}{f_b},
\end{equation}
where $C$ is the number of counts in a full-energy peak,
$\epsilon_{\gamma}$ is the detector energy-dependent full-energy efficiency for the relevant target-detector geometry
and $I_{\gamma}$ is the $\gamma$-intensity per decay.
The $I_{\gamma}$ used in this work was taken from \cite{Datasheet198, GUR16, ABR10}.
The correction due to $\gamma$-ray self absorption in the sample is $K_{\gamma}$.
In the case of a disk sample of thickness $x$, $K_{\gamma}\approx\frac{1-e^{-\mu x}}{\mu x}$,
where $\mu$ is the $\gamma$-ray absorption coefficient.
The $\gamma$-ray absorption coefficients, $\mu$, were taken from \cite{NIST}.
The decay constant of the activated nucleus is $\lambda=\frac{ln(2)}{t_{1/2}}$.
The cooling time between the end of the irradiation and the start of activity measurement is $t_{cool}$,
and $t_{real}$ ($t_{live}$) is the real (live) measurement time.
The decay of activated nuclei during the irradiation is accounted for in $f_b$.
It is calculated using the time dependence of the neutron yield $\Phi(t)$, obtained from the fission chamber (see Fig. \ref{fig: current_Ga}), by
$f_{b}=\frac{\intop_{0}^{t_{a}}\Phi(t)e^{-\lambda(t_{a}-t)}dt}{\intop_{0}^{t_{a}}\Phi(t)dt}$. $t_a$ is the time of the end of irradiation.
The decay parameters and correction factors used in this analysis are listed in Tables \ref{table: decay_Ga} and \ref{table:ana_data}.
\begin{table*}[htpb]
\caption{Analysis of one of the $\gamma$-ray spectra from the $^{nat}$Ga$_2$O$_3$ \#2 sample,
measured 5180 s after the end of the neutron irradiation, with 656.8 s real 
time and 600 s live time; see Eq. (\ref{eq:N_act_exp}) for the definition of notations.
The data for $E_{\gamma}$ and $I_{\gamma}$ were taken from \cite{GUR16, ABR10}.}
\label{table:ana_data}
\begin{ruledtabular}
\ra{1.3}
\begin{tabular}{l c c c c c c c}
Nucleus & $E_{\gamma}$ (keV) & counts &$I_{\gamma}$ (\%) & $\epsilon_{\gamma}$ ($10^{-4}$) & $K_{\gamma}$ & $f_b$ & $N_{act}$ ($10^9$) \\
[0.5ex]
\hline
$^{70}$Ga & 176.115(13) & 1264(93) & 0.29(1) & 30.4(8) & 0.970(3) & 0.550 & 16.4(14)\\
& 1039.513(10) & 602(43) & 0.65(5) & 6.50(8) & 0.990(3) & 0.550 & 16.1(17)\\
$^{72}$Ga & 600.912(15) & 3578(99) & 5.822(19) & 10.5(1) & 0.987(3) & 0.984 & 7.96(24)\\
& 629.967(19) & 15484(157) & 26.13(4) & 10.0(1) & 0.987(3) & 0.984 & 8.00(12)\\
& 834.13(4) & 45243(224) & 95.45(8) & 7.87(9) & 0.989(3) & 0.984 & 8.15(10)\\
& 894.327(18) & 4419(80) & 10.136(15) & 7.41(9) & 0.989(3) & 0.984 & 7.96(17)\\
& 1050.794(17) & 2643(64) & 6.991(11) & 6.44(8) & 0.990(3) & 0.984 & 7.93(22) \\
& 1861.996(18) & 1248(46) & 5.41(3) & 3.92(8) & 0.992(3) & 0.984 & 7.93(17)\\
& 2201.586(17) & 5479(79) & 26.87(12) & 3.39(7) & 0.993(3) & 0.984 & 8.10(21)\\
& 2491.026(17) & 1437(42) & 7.73(3) & 3.05(6) & 0.993(3) & 0.984 & 8.22(29)\\
& 2507.718(17) & 2454(52) & 13.33(6) & 3.03(6) & 0.993(3) & 0.984 & 8.19(25)\\ [0.5ex]
\end{tabular}
\end{ruledtabular}
\end{table*}
The numbers of activated nuclei at the end of the irradiation, $N_{act}$, calculated with Eq. (\ref{eq:N_act_exp}), are summarized in table \ref{table: Nact summary_Ga}.
\begin{table}[htpb]
\centering
\caption{\label{table: Nact summary_Ga}The number of activated nuclei at the end of the irradiation, $N_{act}$, and comparison with simulated $N_{act}$.}
\begin{ruledtabular}
\ra{1.3}
\begin{tabular}{l c c c}
Sample& Nucleus & $N_{act}$ ($10^9$) & simulated $N_{act}$ ($10^9$)\\ [0.5ex]
\hline
Au-14& $^{198}$Au& 4.07(8) & 4.09\\
$^{nat}$Ga$_2$O$_3$ \#2 & $^{70}$Ga & 16.3(7)\\
& $^{72}$Ga & 8.1(1)\\
Au-15& $^{198}$Au& 3.65(6) & 3.63\\ [1ex]
\end{tabular}
\end{ruledtabular}
\end{table}


\section{Experimental cross section \label{sec:cs}}
\begin{figure}[htbp]
\centering
 \includegraphics[width=\columnwidth]{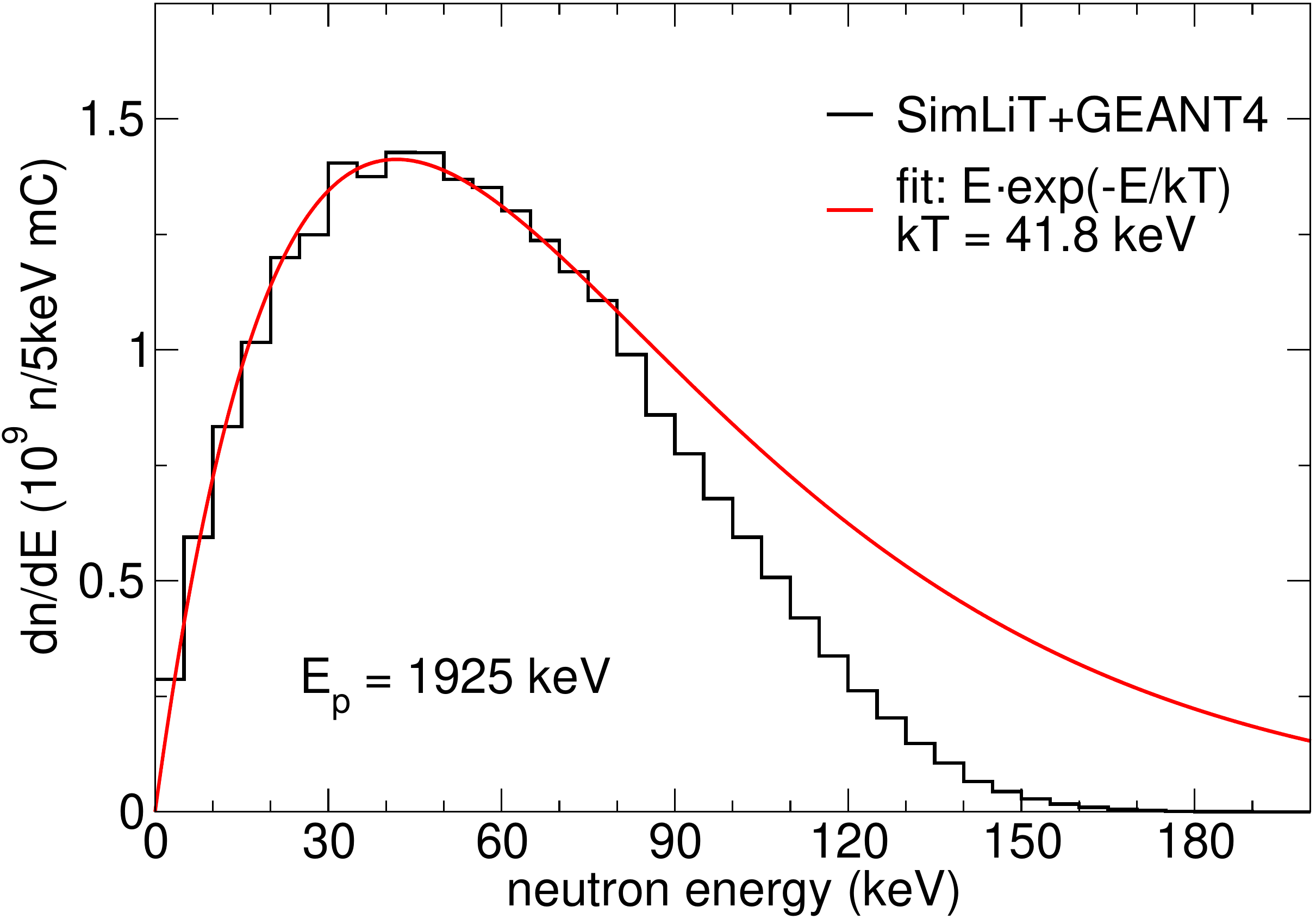}
 \caption{\label{fig: Ga_energy} The simulated neutron spectrum impinging on the Ga target (black).
 Also shown is a fit (between 0 and 90 keV) to a Maxwell-Boltzmann (MB) flux (red), where the best fit is for $kT$ = 41.8 keV.}
 \end{figure}

Since the sample cross section is measured relative to the Au cross section (which is considered known),
the cross section of the sample, averaged over the experimental neutron spectrum, can be expressed as
\begin{equation}
 \sigma_{exp}(i)=\sigma_{ENDF}(\textrm{Au})\frac{N_{act}(i)}{N_{act}(\textrm{Au})}\frac{n_{t}(\textrm{Au})}{n_{t}(i)}, \label{eq:sigma_i}
\end{equation}
where $i$ denotes the Ga stable isotope (69 or 71)
and $\sigma_{ENDF}(\textrm{Au})$ is the reference Au cross section from the ENDF/B-VIII.0 library \cite{ENDF8}
averaged over the experimental neutron spectrum.
It is defined as
\begin{equation}
 \sigma_{ENDF}(\textrm{Au})=\frac{\int\sigma_{ENDF}(E_n; \textrm{Au})\frac{dn}{dE_n}dE_n}{\int\frac{dn}{dE_n}dE_n}. \label{eq:sigma_endf}
\end{equation}
The energy-dependent $^{197}$Au$(n,\gamma)^{198}$Au cross section $\sigma_{ENDF}(E_n; \textrm{Au})$ was taken from ENDF/B-VIII.0 \cite{ENDF8}, in 
agreement with high-precision experimental data \cite{Lederer_Au, Massimi_Au}.
The neutron spectrum, $\frac{dn}{dE_n}$, is obtained from our simulation code (Fig. \ref{fig: Ga_energy}),
developed and benchmarked by experiment \cite{Gitai, SimLiT, PLB_Zr, Paul2019}.
The simulated neutron spectrum impinging on the Ga target, along with a $kT$ = 41.8 keV 
fit to a Maxwellian neutron flux ($\propto E\cdot e^{-\frac{E}{kT}}$), is presented in Fig. \ref{fig: Ga_energy}.
This spectrum is generated by a GEANT4 \cite{GEANT4} simulation, using the SimLiT code \cite{SimLiT} output as the neutron source.
The SimLiT calculation uses $^7$Li$(p,n)$ differential cross sections taken from \cite{LIS75}
and takes into account the proton mean beam energy and energy spread,
proton energy loss in the liquid Li using differential $dE/dx$ values taken from SRIM \cite{ZIE10}
and a Gaussian proton beam profile consistent with the Au monitor auto-radiography (Fig. \ref{fig: Au_radiography}).
The detailed GEANT4 simulation 
takes into account the LiLiT geometry setup including the off-center position of the
neutron beam relative to the Au-Ga-Au target and the surrounding materials.
The simulation explicitly calculates (see \cite{PLB_Zr, Paul2019} for details) the number of  activated $^{198}$Au nuclei,
based on the $\sigma_{ENDF}(E_n;Au)$ cross sections and the measured proton charge 
during irradiation and reproduces the experimental $^{198}$Au activity of the Au monitors within 0.5\% (Table \ref{table: Nact summary_Ga}).

The results of the experimental cross sections, $\sigma_{exp}$, of $^{69}$Ga and $^{71}$Ga (Eq. \ref{eq:sigma_i})
and of the $\sigma_{ENDF}(\textrm{Au})$ values (Eq. \ref{eq:sigma_endf}) are presented
in Table \ref{table: CS summary_Ga}.

\begin{table}[htbp]
\centering
\caption{\label{table: CS summary_Ga}The experimental cross sections, $\sigma_{exp}$ (Eq. \ref{eq:sigma_i}),
measured in this work for the $^{69,71}$Ga$(n,\gamma)^{70,72}$Ga reactions, together with 
the $\sigma_{ENDF}$ cross section averaged over the neutron energy distribution (Fig. \ref{fig: Ga_energy}). See Table \ref{table: uncertainties_Ga} for explanation of $\sigma_{ENDF}(^{197}$Au) uncertainty.}
\begin{ruledtabular}
\ra{1.3}
\begin{tabular}{l c c}
Isotope & $\sigma_{exp}$ (mb) & $\sigma_{ENDF}$ (mb)\\ [0.5ex]
\hline
$^{197}$Au &  &$524(10)$ \\
$^{69}$Ga & $119(5)$ & 103\\
$^{71}$Ga & $89(2)$ & 107\\ [1ex]
\end{tabular}
\end{ruledtabular}
\end{table}

\section{Maxwellian Averaged Cross Section \label{sec:macs}}
The Maxwellian Averaged Cross Section (MACS) at a given thermal energy, $kT$, is defined as:
\begin{equation}
\Braket{\sigma}_{kT}=
\frac{\Braket{\sigma v}}{v_{T}}=\frac{2}{\sqrt{\pi}}
\frac{\intop_{0}^{\infty}\sigma(E_{n})E_{n}e^{-\frac{E_{n}}{kT}}dE_{n}}{\intop_{0}^{\infty}E_{n}e^{-\frac{E_{n}}{kT}}dE_{n}},
\label{eq:MACS}
\end{equation}
where $\sigma(E_{n})$ is the differential $(n,\gamma)$ reaction cross section at neutron energy $E_n$.
In this work, the MACS at a given thermal energy $kT$ is calculated with the procedure developed in \cite{PLB_Zr,Ar_PRL,Paul2019} and using Eq. (\ref{eq:MACS_exp}),
\begin{equation}
\label{eq:MACS_exp}
   \textrm{MACS}_{exp}(kT) = \frac{2}{\sqrt\pi}\cdot C_{lib}(kT) \cdot \sigma_{exp},
\end{equation}
where the correction factor $C_{lib}(kT)$ is given by Eq. (\ref{eq:C_lib}):
\begin{equation}
\label{eq:C_lib}
C_{lib}(kT) =\frac{ \frac{\int_0 ^{\infty} \sigma_{lib}(E_n) E_n e^{-\frac{E_n}{kT}}dE_n}{\int_0 ^{\infty} E_n e^{-\frac{E_n}{kT}}dE_n}} { \frac{\int_0 ^{\infty} \sigma_{lib}(E_n) \frac{dn}{dE_n} dE_n}{\int_0 ^{\infty} \frac{dn}{dE_n} dE_n}}. 
\end{equation}
In Eq. (\ref{eq:C_lib}), $\frac{dn}{dE_n}$ is the simulated experimental neutron spectrum (Fig. \ref{fig: Ga_energy}) and  $\sigma_{lib}(E_n)$ is the energy-dependent neutron capture cross section taken from an evaluation library.

In Table \ref{table: MACS_comp_Ga} we present the correction factors, $C_{lib}$, and the MACS$_{exp}$ at $kT = 30$ keV,
calculated with Eq. (\ref{eq:MACS_exp}), for the Ga isotopes.
The MACS$_{exp}$ at $kT = 30$ keV derived in this work are obtained by using in Eq. (\ref{eq:MACS_exp})
the average $C_{lib}$ calculated using
the various cross section libraries: ENDF/B-VIII.0 \cite{ENDF8},
JENDL-4.0 \cite{Jendl}, JEFF-3.3 \cite{Jeff}, CENDL-3.2 \cite{Cendl32} and TENDL-2019 \cite{TENDL19} (see Table \ref{table: MACS_comp_Ga}).

\begin{table}[htbp]
\centering
\caption{\label{table: MACS_comp_Ga} Comparison of correction factors, $C_{lib}$, and 
MACS calculated for $^{69}$Ga and $^{71}$Ga at 30 keV using the different libraries \cite{ENDF8, Jendl, Jeff, Cendl32, TENDL19}.
See text for explanation.}
\begin{ruledtabular}
\ra{1.3}
\begin{tabular}{l c c c c c}
& \multicolumn{2}{c}{$C_{lib}$} && \multicolumn{2}{c}{MACS$_{exp}$ (mb)}\\
\cmidrule {2-3}
\cmidrule {5-6}
Library& $^{69}$Ga & $^{71}$Ga && $^{69}$Ga & $^{71}$Ga\\ [0.5ex]
\hline
ENDF/B-VIII.0 \cite{ENDF8} & 1.03 & 1.02 && 138  & 103\\
JENDL-4.0 \cite{Jendl} & 1.03 & 1.01 && 138  &  102\\
JEFF-3.3 \cite{Jeff} & 1.0 & 0.97 && 134 & 98\\
CENDL-3.2 \cite{Cendl32} & 0.99 & 1.08 && 133 & 108\\
TENDL-2019 \cite{TENDL19} & 1.04 & 1.15 && 139  & 115\\
\hline
average & 1.018 & 1.046 && 136.4  & 105.0\\
standard deviation & 0.02 & 0.07 && 2.8 & 6.7\\[1ex]
\end{tabular}
\end{ruledtabular}
\end{table}

The experimental uncertainties, as discussed in detail in \cite{PLB_Zr, Paul2019}, are summarized in Table \ref{table: uncertainties_Ga}.
The uncertainties for the average $C_{lib}$ were obtained by taking the standard deviation
of the $C_{lib}$ calculated for the various cross section libraries \cite{ENDF8, Jendl, Cendl32, Jeff, TENDL19} (see Table \ref{table: MACS_comp_Ga}).

\begin{table}[htp]
\centering
\caption{\label{table: uncertainties_Ga}Random (rand) and systematic (sys) uncertainties in the results presented in this work.}
\begin{ruledtabular}
\begin{threeparttable}
\ra{1.3}
\begin{tabular}{l c c c c c c c}
 & \multicolumn{5}{c}{Uncertainty (\%)}\\
\cmidrule{2-6}
 & \multicolumn{2}{c}{$^{69}$Ga} && \multicolumn{2}{c}{$^{71}$Ga}\\
 Source of uncertainty&rand&sys&&rand&sys\\ [0.5ex]
 \hline
target thickness& 0.5&&& 0.5\\
activity measurement & 4.0  &&& 0.5   \\
full-energy eff. rel. to Au && 0.5\tnote{a} &&& 0.5\tnote{a}\\
intensity per decay && 3.4 &&& 0.1\\
$\sigma_{ENDF}$(Au)&&1.9\tnote{b}&&&1.9\tnote{b} \\
$C_{lib}$ && 2.0  &&& 6.4  \\
\hline
Total random uncertainty& 4.0 &&& 0.7 \\
Total systematic uncertainty&& 4.4 &&& 6.7 \\
\hline
Total uncertainty &\multicolumn{2}{c}{6.0} && \multicolumn{2}{c}{6.7}\\[1ex]
\end{tabular}
\begin{tablenotes}
\item[a] This contribution to the uncertainty of the $^{69,71}$Ga MACS includes only the ratio of the full-energy efficiencies of the $^{69,71}$Ga $\gamma$ lines to that of the $^{198}$Au $\gamma$ line. In Tables \ref{table:ana_data} and \ref{table: Nact summary_Ga} the overall uncertainty (including the systematic uncertainty of the calibration sources) is quoted.
\item[b] This value includes the uncertainty of beam parameters (proton beam energy, energy spread, and distance of sample from Li) of 0.6\%,
the uncertainty of the simulations of 1.5\% and the uncertainty of the ENDF cross section for Au of 1.0\%, $\sqrt{0.6^2+1.5^2+1.0^2} = 1.9\%$.
See \cite{Paul2019} for more details of the uncertainties.
\end{tablenotes}

\end{threeparttable}
\end{ruledtabular}
\end{table}


\begin{table*}[htpb]
\centering
\caption{\label{table:SACS_MACS}
 Comparison of the results of this work with previous experiments
 and compilations.
 The MACS listed in the Table correspond to $kT$= 30 keV.
 The previously measured MACS were renormalized as specified in 
 \cite{kadonis1} using the recent measured 
 $^{197}$Au$(n,\gamma)^{198}$Au cross section data \cite{Massimi_Au, Lederer_Au, MAS10} used as standard.
 The KADoNiS v0.3 \cite{Kadonis} recommended MACS (30 keV) is the 
 average of the experimental data renormalized to the 
 $^{197}$Au$(n,\gamma)^{198}$Au MACS data measured in \cite{LiBe}.
The KADoNiS v1.0 \cite{kadonis1} recommended MACS(30 keV) values are an average from evaluated libraries ENDF/B-VII.1 \cite{ENDF}, JENDL-4.0 \cite{Jendl} and TENDL-2015 \cite{TENDL15}.}
\begin{ruledtabular}
\begin{threeparttable}
\centering
\ra{1.3}
\begin{tabular}{l c c c c c c c}
& KADoNiS & KADoNiS & Walter, & Anand & Walter & G{\"o}bel & This work\\
&v0.3 \cite{Kadonis} & v1.0 \cite{kadonis1} &
1984 \cite{WAL84} & \textit{et al.}, 1979 \cite{ANA79} & \textit{et al.}, 1986 \cite{WAL86} &  \textit{et al.}, 2021 \cite{GOB21} \\ [0.5ex]
\hline
MACS($^{69}$Ga) (mb) & 139(6) & 123(9) & 149(6) & & & & 136(8)\\
$\sigma_{exp}$($^{69}$Ga)/$\sigma_{exp}$($^{197}$Au) \tnote{1} & & & & & & 0.286(19) & 0.227(12) \\
\\
MACS($^{71}$Ga) (mb) & 123(8) & 103(14) & & 79(23) & 130(8)& & 105(7)\\
$\sigma_{exp}$($^{71}$Ga)/$\sigma_{exp}$($^{197}$Au) \tnote{1} & & & & & & 0.173(11) & 0.170(5)\\ [1ex]
\end{tabular}
\begin{tablenotes}
    \item[1] Note that G{\"o}bel \textit{et al.} \cite{GOB21} denote
    $\sigma_{exp}$ as defined in Eq. (\ref{eq:sigma_i}) by SACS (spectrum averaged cross section) for their experimental neutron spectrum. 
 \end{tablenotes}
\end{threeparttable}
\end{ruledtabular}
\end{table*}

\begin{figure}[htpb]
\centering
 \includegraphics[width=\columnwidth]{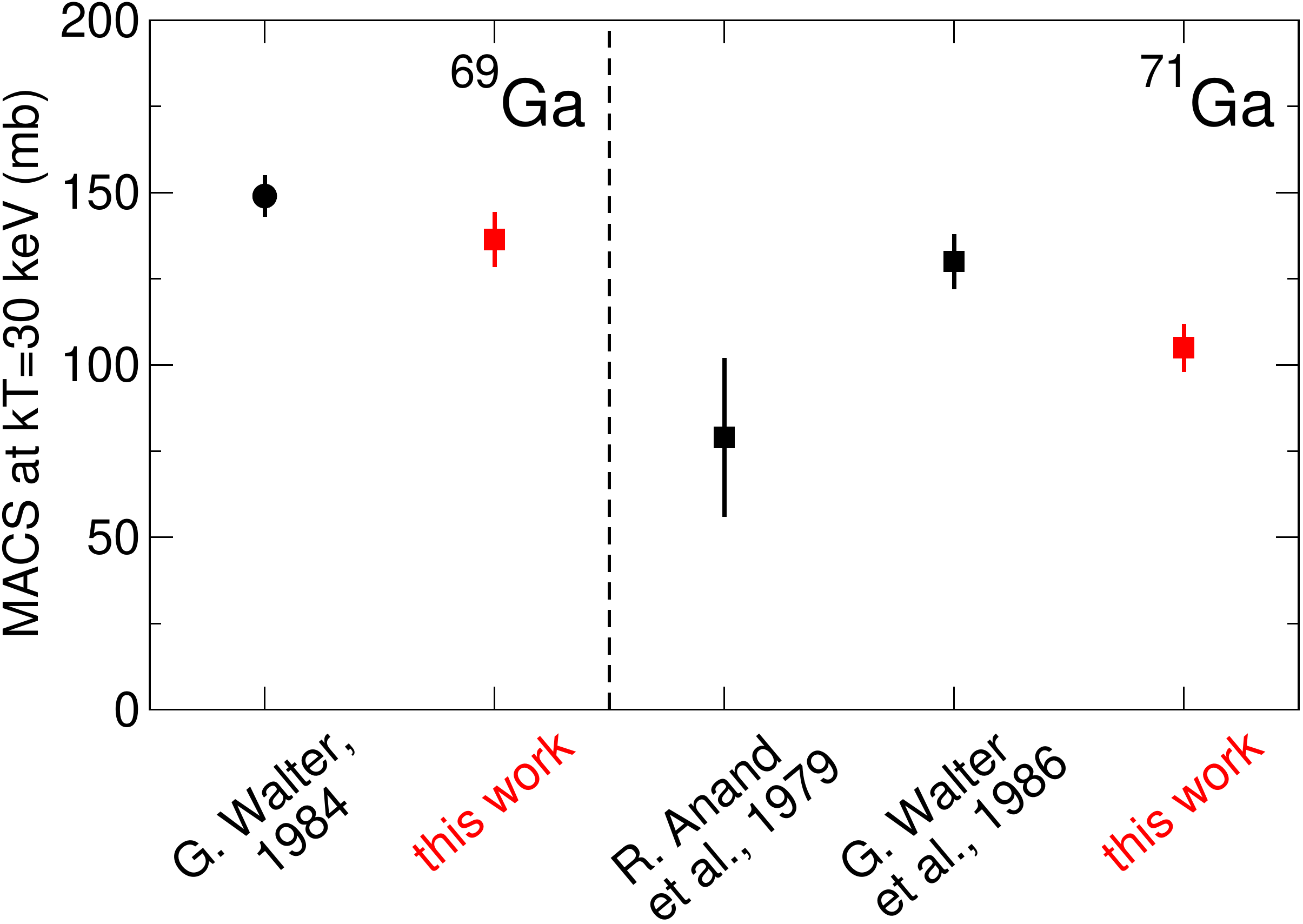}
 \caption{\label{fig: MACS_comp} Comparison of this work's results (red) with previously measured MACS at $kT=30$ keV (black)
 for $^{69}$Ga \cite{WAL84} (left) and $^{71}$Ga \cite{ANA79, WAL86} (right).
 The previously measured MACS were renormalized as specified in the KADoNiS v1.0 website \cite{kadonis1}.
 The TOF measurement is displayed as a circle and the activation measurements as squares.}
 \end{figure}

\section{Discussion \label{sec:disc}}
This work's results are compared in Table \ref{table:SACS_MACS} and Fig. \ref{fig: MACS_comp}
with previously measured MACS at $kT$ = 30 keV
for $^{69}$Ga \cite{WAL84} and $^{71}$Ga \cite{ANA79,WAL86}, and with
compilations of experimental values (KADoNiS v0.3, \cite{Kadonis}) and of evaluated values (KADoNiS v1.0, \cite{kadonis1}).
We compare also the ratios $\sigma_{exp}$($^A$Ga)/$\sigma_{exp}$($^{197}$Au)
with the corresponding values from G{\"o}bel \textit{et al.} \cite{GOB21}, denoted there as SACS($^A$Ga)/SACS($^{197}$Au),
calculated for the relevant experimental spectrum.

Our result for the $^{69}$Ga MACS (30 keV) are in good agreement within uncertainties with the previous experimental value of
\cite{WAL84} and those recommended by Bao \textit{et al.} \cite{BAO00} and KADoNiS v0.3 \cite{Kadonis}).
Our MACS value for $^{71}$Ga is larger than the experimental value of \cite{ANA79}, but smaller than the experimental value of \cite{WAL86}
recommended by Bao \textit{et al.} \cite{BAO00}
and KADoNiS v0.3 \cite{Kadonis}), though marginally consistent within the quoted uncertainties.
This smaller value is significant in view of the fact that the previous value of KADoNiS v0.3 \cite{Kadonis} was used in extensive network calculations \cite{PIG10}; see below.
Both $^{69,71}$Ga MACS values extracted in this work are in reasonable agreement with the evaluated values of KADoNiS v1.0 \cite{kadonis1}.
We note also the good agreement of our
$\sigma_{exp}$($^{71}$Ga)/$\sigma_{exp}$($^{197}$Au)
 value with that of G{\"o}bel \textit{at al.} \cite{GOB21},
 but a striking discrepancy for
 $\sigma_{exp}$($^{69}$Ga)/$\sigma_{exp}$($^{197}$Au),
 whose origin is not understood.

Gallium, like the other elements between iron and strontium (26 $<$ Z $<$ 38, 60 $\lesssim$ A $\lesssim$ 90),
is produced primarily by the weak component of the $s$ process in massive stars (M$_{initial}>$ 8M$_\odot$)
(\cite{KAP89} and references therein). Pignatari \textit{et al.} \cite{PIG10} studied nucleosynthesis,
including the weak $s$ process, in a model of a Population I (solar metallicity) 25M$_\odot$ star.
For the $^{69,71}$Ga$(n,\gamma)$ MACS, the authors used the values recommended by Bao \textit{et al.} \cite{BAO00} (KADoNiS 0.3 \cite{Kadonis}).
As shown in Table \ref{table:SACS_MACS} and discussed above, the new $^{69}$Ga MACS measurement presented here supports the Bao \textit{et al.} \cite{BAO00} value,
but the $^{71}$Ga MACS is 15\% lower.
This means that less $^{71}$Ga is consumed by the $(n,\gamma)$ reaction and hence its 
isotopic fraction is expected to increase compared to that calculated by Pignatari \textit{et al.} \cite{PIG10}.
In addition, since the $s$ process flow goes from Ga to Ge,
$^{69}$Ga$(n,\gamma)^{70}$Ga$(\beta^-)^{70}$Ge and
$^{71}$Ga$(n,\gamma)^{72}$Ga$(\beta^-)^{72}$Ge (see Fig. \ref{fig:Ga_s_flow}), the lower $^{71}$Ga MACS is expected to
result in a lower production of $^{72,73,74}$Ge and reduce their isotopic fraction.
The lower $^{71}$Ga MACS (105(7) mb) may have as well a wider effect, as Pignatari \textit{et al.} \cite{PIG10}
concluded that the effects of MACS $\lesssim 150$ mb tend to propagate to heavier isotopes.
However, we should caution that the above potential effects are tentative.
The only way to check for the effect of the new $^{71}$Ga MACS is to incorporate it in a network calculation like the one performed by Pignatari \textit{et al.} \cite{PIG10}.

Such nucleosynthesis models
need to be tested against isotopic compositions in the relevant star types.
This is done for many elements by isotopic analysis of chemical elements in presolar grains. However, to date, there are no measurements of Ga in presolar grains (\cite{HYN09,STE20,Presolar}).
The most studied family of presolar grains are the carbides, mainly SiC and graphite.
At the same time, Ga is thought to usually not form carbides \cite{KUM17},
so it is unlikely that there is enough Ga in presolar carbides for isotopic analysis
with the current capabilities of the experimental techniques used.
On the other hand, Lodders \cite{Lodders} calculated that Ga condensed in the early
Solar system as a trace element into the mineral feldspar,
by substituting for the major element Al in the crystal lattice,
and into Fe metal to a lesser extent.
While presolar feldspar has not been found to date, most of the presolar oxides and silicates studied to date contain Al as a major element.
This makes them more likely hosts for stellar Ga to be analyzed.
 
\section{Summary}
The neutron capture cross sections of $^{69,71}$Ga were measured
by the activation technique in the intense $kT \approx$ 40 keV quasi-Maxwellian neutron field of the SARAF-LiLiT facility.
The reaction products were measured by $\gamma$ spectrometry with a HPGe detector.
The experimental cross sections were converted to Maxwell-averaged cross sections
at $kT$ = 30 keV
using the energy dependence of various neutron cross section library data.
The MACS values obtained in this work are MACS ($^{69}$Ga) = 136(8) mb and MACS ($^{69}$Ga) = 105(7) mb.
The $^{69}$Ga MACS value is in good agreement with previous
experimental values and their recommended values while that of $^{71}$Ga is smaller and with reduced uncertainty than the
experimental recommended value.
This smaller value may have implications in network calculations such as those of Pignatari \textit{et al.} \cite{PIG10} which used so far the recommended values.
Potential natural samples to measure stellar Ga composition are presolar silicates and oxides,
in which Ga may replace Al in the grains' crystal lattice.
We note a significant and not understood discrepancy between our
experimental cross-section value of $^{69}$Ga with that recently published by G{\"o}bel \textit{et al.} \cite{GOB21},
beyond that expected from the different experimental conditions.

\begin{acknowledgments}
We would like to thank the SARAF and LiLiT (Soreq NRC) staffs for their dedicated help during the experiments.
This work was supported in part by Israel Science Foundation (Grant Nr. 1387/15) and the Pazy Foundation (Israel).
M.P. acknowledges support by the European Union (ChETEC-INFRA, project
no. 101008324).

\end{acknowledgments}

%

\end{document}